\documentclass[useAMS,usenatbib,usegraphicx]{mn2e}
\usepackage{amsmath,amsfonts,amssymb}

\title [The rest-frame K-Band Luminosity Function]{The effect of
  thermally pulsating asymptotic giant branch stars on the evolution
  of the rest-frame near-infrared galaxy luminosity function}

\author[Henriques et al.]
{Bruno Henriques$^{1}$\thanks{E-mail: Bruno.Henriques@port.ac.uk}, 
Claudia Maraston$^{1}$, Pierluigi Monaco$^{2,3}$,
\newauthor
Fabio Fontanot$^{2}$,
Nicola Menci$^{4}$, Gabriella De Lucia$^{2}$, Chiara Tonini$^{1}$\\
  {}$^{1}$Institute of Cosmology and Gravitation, University of
  Portsmouth, Portsmouth PO1 3FX, United Kingdom\\
  {}$^{2}$INAF - Astronomical Observatory of Trieste, via G.B. Tiepolo 11, I-34143 Trieste, Italy\\
  {}$^{3}$Dipartimento di Fisica - Sezione di Astronomia, Universita` di Trieste, via G.B. Tiepolo 11, I-34131 Trieste, Italy\\
  {}$^{4}$INAF - Osservatorio Astronomico di Roma, via di Frascati 33, I-00040 Monteporzio, Italy}

\begin{document}

\date{Submitted to MNRAS}

\pagerange{\pageref{firstpage}--\pageref{lastpage}} \pubyear{2009}

\maketitle

\label{firstpage}

\begin{abstract}

  We address the fundamental question of matching the rest-frame
  $K$-band luminosity function (LF) of galaxies over the Hubble time
  using semi-analytic models, after modification of the stellar
  population modelling. We include the Maraston evolutionary synthesis models, that
  feature a higher contribution by the Thermally Pulsating -
  Asymptotic Giant Branch (TP-AGB) stellar phase, into three different
  semi-analytic models, namely the De Lucia and Blaizot version of the
  Munich model, {\sc morgana} and the Menci model. We leave all other
  input physics and parameters unchanged. 

  
  We find that the modification of the stellar population emission can
  solve the mismatch between models and the observed rest-frame
  $K$-band luminosity from the brightest galaxies derived from UKIDSS
  data at high redshift. For all explored semi-analytic models this
  holds at the redshifts - between 2 and 3 - where the discrepancy was
  recently pointed out. The reason for the success is that at these
  cosmic epochs the model galaxies have the right age ($\sim$1 Gyr) to
  contain a well-developed TP-AGB phase which makes them redder
  without the need of changing their mass or age. We have also
  computed a version of the Munich model using the Charlot and Bruzual
  models that adopt the Marigo TP-AGB prescription and find the same
  result as with the Maraston models.

  At the same time, the known overestimation of the faint end is
  enhanced in the $K$-band when including the TP-AGB contribution. At
  lower redshifts ($z<2$) some of the explored models deviate from the
  data. This is due to too short merging timescales and inefficient
  'radio-mode' AGN feedback. Our results show that a strong
  evolution in mass predicted by hierarchical models is compatible
  with no evolution on the bright-end of the $K$-band LF from z=3 to
  the local universe. This means that, at high redshifts and
    contrary to what is commonly accepted, $K$-band emission is not
    necessarily a good tracer of galaxy mass.

\end{abstract}


\section{Introduction}
\label{sec:intro}

Models for the formation and evolution of galaxies in a Cold Dark
Matter universe (e.g. the so-called semi-analytic models) predict the
intrinsic properties of galaxies, such as ages, metallicities, stellar
masses, star formation rates, etc.., after having tuned a number of
free parameters that make up for the poorly known aspects of baryonic
physics \citep[see][for an extensive review]{Baugh2006}.  The
comparison between models and observations helps constraining these
parameters and robust statistical tools have been recently used to
achieve this goal \citep{Kampakoglou2008, Henriques2009,
  Henriques2010, Bower2010, Lu2010}.

The results of these comparisons are very sensitive to the
spectro-modelling of the stellar component. Either for extracting
galaxy properties such as mass, age, star formation rate from data,
and compare them to the intrinsic quantities of the semi-analytic
models, or for calculating the spectra of semi-analytic galaxies and
compare it with the observed light, the details on how the stellar
modelling is performed influence the final result.

In order to obtain the spectral energy distribution (SED), or specific
broad-band luminosities, of a model galaxy of given mass and star
formation history, evolutionary populations synthesis (EPS) models
(e.g. \citealt{Tinsley1972}, \citealt{Bruzual1983},
\citealt{Buzzoni1989}, \citealt{Bruzual1993}, \citealt{Worthey1994},
\citealt{Vazdekis1996}, \citealt{Fioc1997}, \citealt{Maraston1998},
\citealt{Leitherer1999}, \citealt{Bruzual2003}, \citealt{Thomas2003},
\citealt{Maraston2005} (M05), \citealt{Conroy2009}) are
adopted.  By relying on stellar evolution theory and model atmosphere
calculations or empirical libraries, EPS models provide the expected
spectral energy distribution of a galaxy of given mass and star
formation history.

In galaxy formation models the unit single burst EPS models or Simple
Stellar Populations (SSPs) are used to model coeval stars with a
homogeneous metallicity, after adopting an Initial Mass Function
(IMF). The total stellar emission from the synthetic galaxy composite
population is then obtained by combining SSPs. Hence, what matters on
the final result in terms of stellar evolution are the properties of
the simple stellar population models.

It is clear then that this modelling is a crucial aspect of galaxy
formation and evolution theory. Uncertainties in the conversion
between masses/ages and light can create artificial discrepancies,
which in turn could drive into difficult attempts to modify the
parametrization of the complicated physics of gas cooling, star
formation or feedback to account for this mismatch. The approach we
take in this paper, following our previous work \citep{Tonini2009,
  Tonini2010, Fontanot2010}, is to check the impact of modifying the
input stellar population model in galaxy formation models following
recent progress in the literature.

At present, the highest source of discrepancy between different SSP
models is the treatment of the Thermally-Pulsating Asymptotic Giant
Branch (TP-AGB) phase (\citealt{Maraston1998}, M05,
\citealt{Marigo2008}, \citealt{Conroy2009}). The TP-AGB phase is the
last luminous phase in the Hertzsprung-Russel diagram before
intermediate-mass stars evolve to their final destiny as planetary
nebulae and white dwarfs. TP-AGB stars are very luminous and
cool. Their emission affects the integrated model spectra at
wavelengths larger than $\sim 6000\AA$, peaking around the
$J,H,K$~bands (M05) and a recent study has highlighted their
importance also long-ward the rest-frame $K$~\citep{Kelson2010}. Due to
difficulties in the stellar modelling of this phase, which in turn are
due to mass-loss and the pulsating regime \citep[see][for a
review]{Iben1983}, generally stellar tracks did not include the full
TP-AGB, so did not stellar population models based on these tracks
(see M05 for discussion). \citet{Maraston1998} and M05 include the
TP-AGB semi-empirically calibrating the theoretical energetics with
Magellanic Cloud clusters, an approach now adopted for including the
TP-AGB phase in isochrones \citep{Marigo2007, Bruzual2007}.


The galaxy formation models we use in this study, in their standard
stellar populations, either neglect or do not include a full
contribution from the TP-AGB stellar phase. As pointed out by several
papers (\citealt{Maraston1998}, \citealt{Maraston2001},
\citealt{Maraston2004}, M05, \citealt{Maraston2006},
\citealt{Wel2006}, \citealt{Marigo2007}, \citealt{Bruzual2007},
\citealt{Eminiam2008}, \citealt{Conroy2009}) the inclusion of TP-AGB
stars provides an enhancement of the near-infrared emission of
galaxies dominated by $\sim$1 Gyr old populations. To test the
influence of this inclusion on model predictions, one needs to
consider a statistically significant sample of model galaxies,
covering both a wide range of K-band luminosities and redshifts. This
test is particularly important when the photometric properties of
high-z (i.e. 2$<$z$<$3) galaxies are considered, since we expect them
to be dominated by young stellar populations.


Semi-analytical techniques represent an obvious tool to perform such
test. Predictions from these models have already been compared with
the evolution of the observed $K$-band Luminosity Function (LF)
\citep{Pozzetti2003, Cimatti2004, Kitzbichler2007,
  Cirasuolo2008}. These works consistently found a lack of bright
sources at high redshift. This apparent mismatch is being referred to
as one of the strongest discrepancies between models and data (in
particular in connection with the evolution of the stellar mass
function, see \citealt{Fontanot2009b}b for a critical review of the
latter issue).  However, these comparisons involved
spectro-photometric codes based on stellar tracks where the full
effect of TP-AGB stars was not taken into account.


The first test of this kind is performed in \citet{Tonini2009,
  Tonini2010}. They run the GALICS semi-analytic model
\citep{Hatton2003} using M05 as input EPS and show that the
optical-to-near-IR colours of z$\sim$2 galaxies can be matched by this
type of models, with the original GALICS, which was based on a
pre-TP-AGB EPS, failing to match the observations by a large margin.
Moreover, \citet{Fontanot2010} introduced the M05 stellar populations
in {\sc morgana} obtaining a good match on the number density of
extremely red objects (EROs) at high redshift.

Here we investigate whether the inclusion of the TP-AGB phase has also
an impact on the inability of semi-analytic models to match the galaxy
rest-frame $K$-band luminosity function at high-redshift, namely the
UKIDSS data from \citet{Cirasuolo2008}. This homogeneous data set
covers an ideal redshift range (0$<$z$<$3) for this test.

To perform this analysis, we use three different semi-analytic models
of galaxy formation: the \citet{Delucia2007} version of the Munich
model, {\sc morgana} \citep{Monaco2007} and the \citet{Menci2006}
semi-analytic model. We compare the predictions obtained for the
properties of the galaxy population using both stellar populations
with and without a full treatment of the TP-AGB phase. For the three
models \citep{Delucia2007, Monaco2007, Menci2006} outputs are produced
using, respectively \citet{Bruzual2003}, \citet{Silva1998} and
\citet{Bruzual1993}, and M05. Note that, despite including the
contribution from TP-AGB stars in their model, \citet{Bruzual2003}
only partially account for the emission during this stellar phase,
meaning that 1 Gyr old populations are roughly only half as bright in
the $K$-band when compared to more recent treatments
\citep{Bruzual2007}


This paper is organized as follows. In Section \ref{sec:models}, we
briefly describe the semi-analytic models used in this study, and we
explain how the M05 EPS is implemented in each galaxy formation
model. In Section \ref{sec:data} we describe the data used for
comparison and clarify where the impact of the M05 models is expected
to be found.  Section \ref{sec:results} presents the results for the
evolution of the rest-frame near-infrared luminosity function and in
Section \ref{sec:conclusions} we summarize our conclusions.

\section{The Semi-Analytic Models}
\label{sec:models}

A clear advantage of the work presented in this paper is the use of
three semi-analytic models developed by independent groups and
implementing different techniques for the description of the physics
controlling galaxy formation and evolution. This allows us not only to
assess the impact of the TP-AGB phase, but also to understand the
interplay between the new ingredient and the other assumptions
regarding galaxy physics on the light of the same stellar evolution
background. In particular we consider the \citet{Delucia2007} version
of the Munich model; {\sc MORGANA} (originally described in
\citealt{Monaco2007} and updated by \citealt{Lofaro2009}); and the
\citet{Menci2006} model.

The backbone for all models is a description of the redshift evolution
of the mass and number density of dark matter halos in terms of their
merger history (the so-called merger trees). The evolution of the
baryonic component hosted by these halos is then followed by means of
an approximated set of simplified formulae, aimed at describing the
physical processes acting on the gas (such as gas cooling, star
formation and feedback) in terms of the physical properties of each
model galaxy and/or its components (i.e the stellar, hot and cold gas
content and distribution). These analytical {\it recipes} include a
set of parameters which are usually calibrated against a well defined
subset of low-redshift observations.

The three models adopt different techniques to describe the dark
matter merger trees\footnote{The Munich model uses merger trees
  extracted from a direct N-Body simulation of a cosmological volume
  (the Millennium Simulation, \citep{Springel2005}), {\sc morgana}
  uses the Lagrangian semi-analytic code {\sc pinocchio}
  \citep{Monaco2002} and \citet{Menci2006} uses Monte Carlo
  realizations of merger trees based on the halo merging probability
  given by the Extended Press-Schechter formalism.} and slightly
different cosmologies. However, we do not expect these to have a
significant effect on our conclusions (see e.g. \citealt{Wang2008}).

  
On the other hand, the different star formation histories and the
corresponding distribution of ages and the mass build up in the models
do matter. In the following, we will briefly account for differences
between the models, focusing in particular on the AGN feedback and the
merging time scales, the processes most relevant for the evolution of
the bright-end of the $K$-band LF. For more details on the treatment
of these physical processes in the different models we refer the
reader to the original papers, and to \citet{Delucia2010}; Fontanot et
al. (2009b) for recent comparisons.

\subsection{AGN Feedback}

The recipe adopted to describe AGN feedback is of crucial importance,
since it largely determines the stellar population properties of the
most massive galaxies, whose evolution is the focus of our
paper. Recent studies \citep[e.g.][]{Croton2006} assume that the
growth of Super-Massive Black-Holes (SMBHs) at the center of model
galaxies follows two channels, a ``bright-mode'' (or ``quasar-mode'')
and a ``radio-mode'', related to the efficient production of radio
jets. The ``quasar-mode'' is fueled by merger driven instabilities, it
is the dominant channel in terms of black hole growth and can be
effective in producing feedback at early times (where merging rates
are high). The ``radio-mode'' is less important in terms of SMBH
growth but is responsible for star formation quenching at low
redshift.

The details of the implementations of the two modes differ between the
models considered in this study (see e.g. \citealt{Fontanot2010b}b for
a detailed discussion about {\sc morgana} and
\citet{Delucia2007}). While the net effect of the ``quasar-mode'' is
quite similar between the various models, even if the implementations
are slightly different, it is the ``radio-mode'', to be mostly
responsible for differences in the galaxy stellar populations towards
low redshift.  

In the Munich model, the ``radio-mode'' feedback is the result of
quiescent gas accretion from a static hot halo \citep{Croton2006},
with no triggering mechanism required.  In {\sc morgana}, the
``radio-mode'' is due to the accretion (at very low rates) of cold gas
from a reservoir surrounding the central SMBH (see
\citet{Fontanot2006} for more details). Note that some amount of star
formation is required to destabilize the gas in the reservoir. Hence,
star formation is not completely quenched. This residual star
formation causes galaxies to have colours that are too blue at low
redshift with respect to both observations and other models
\citep{Kimm2009} and contributes to an excessive build up of massive
objects at later times. Finally, the \citet{Menci2006} model does not
include ``radio-mode'' feedback. For this reason, at low redshift,
massive objects always have ongoing star formation, which causes an
excessive mass build up in these objects.  Relevant to our work is
that this results in an over-prediction of the bright tail of the
K-band LF, as we will show in Section \ref{sec:results}.

\subsection{Merging Times}

Dark matter substructures and their clustering have relevant
consequences on the evolution of galaxies. Gravitational processes
such as dynamical friction and tidal stripping affect the morphology,
the stellar and the gaseous content of galaxies. Two-body mergers are
even more extreme processes, leading to the formation of a new object,
whose final properties depend on the properties of the progenitors.

In the Munich model, dark matter substructures in the N-body
simulation are explicitly tracked down until tidal truncation and
stripping reduce their mass below the resolution limit of the
simulation \citep{Delucia2004a,Gao2004}. In \citet{Menci2006} dark
matter is followed using a Press-Schechter formalism and satellite
halos are partially disrupted as the density in their outer parts
becomes lower than the density of the host halo within the pericentre
of its orbit (see \citet{Menci2002} for details). After this point the
merging time of the satellite in both models is computed using the
classical \citet{Chandrasekhar1943} dynamical friction
approximation. It is worth stressing that the \citet{Delucia2007}
model includes an additional parameter in this formula, effectively
doubling the expected merging times. This value was introduced to
reduce the slight excess of bright galaxies that would be produced
otherwise.
{\sc morgana} does not track explicitly dark matter substructures and
assumes that satellite galaxies merge onto central galaxies after a
dynamical friction time-scale which is computed using analytic
formulae proposed by \citet{Taffoni2003}.

\citet{Delucia2010} compare different approximations for the dynamical
friction merging time-scales used in semi-analytics. They find that
while the \citet{Delucia2007} recipe is in good agreement with some
recent results based on $N$-body-simulations
\citep{BoylanKolchin2008}, the \citet{Taffoni2003} formulae predict
significantly shorter merging times. Note that the same is true for
\citet{Menci2006} with merging times two times shorter than in
\citet{Delucia2007}.

Despite the overall agreement between different models in terms of the
mass build up found by Fontanot et al. (2009b), it can be seen that
{\sc morgana} shows an excessive build up of massive galaxies at late
times. We expect \citet{Menci2006} to show a similar behaviour. This
is due to the combined effect from the enhanced merger activity and
the ongoing star formation due to inefficient AGN feedback at low
redshift. This will affect our results with both models
over-estimating the number density of bright $K$-band objects at later
times (see Section \ref{sec:results}).
  
\subsection{Implementation of the M05 Models}
\label{sub:implementationm05}

As recalled in the Introduction, the spectra of galaxies in
semi-analytic models are obtained by means of spectro-photometric
population synthesis models. The implementation of the M05 models is
straightforward in these semi-analytics.  We use SSPs corresponding to
four metallicities, 1/20 $Z_{\odot}$, 1/2 $Z_{\odot}$, $Z_{\odot}$ and
2 $Z_{\odot}$, which despite not being exactly the same as for the
stellar populations previously used (since the input stellar tracks of
the M05 models are different, see M05 for details), cover a similar
range and are as coarse. Therefore, this difference has no impact on
our predictions. The same IMF that was previously adopted in the
various semi-analytic models is retained, namely the
\citet{Chabrier2003} for {\sc morgana}\footnote{Note that the M05
  stellar populations were implemented in {\sc morgana} by
  \citet{Fontanot2010}} and the Munich model, and a
\citet{Salpeter1955} for \citet{Menci2006}.

The predicted luminosities are then corrected for dust extinction.
For all models we keep these prescriptions unchanged.
%
The different treatment of dust extinction has non-negligible effects
on the predicted magnitudes and colors, especially at $z>2$
\citep{Fontanot2009}. However, since the rest-frame $K$-band emission is
relatively insensitive to dust attenuation, we do not expect these
differences to substantially affect our results.

Finally, It is worth stressing that we keep all other assumptions and
parameters of the semi-analytic models as in their original
formulation. Therefore, we can highlight any modification due just to
the change in the stellar population libraries.

\begin{figure}
\centering
\includegraphics[width=8.4cm]{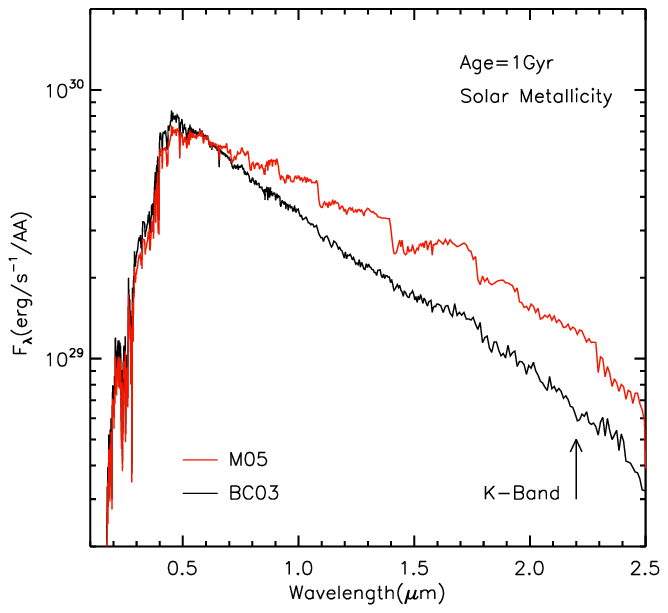}
\caption{A simple stellar population spectral energy distribution from
  M05 is compared with the equivalent predictions from models used in
  semi-analytic models (here \citet{Bruzual2003} as an illustrative
  example).  The plot refers to a 1 Gyr old population with solar
  metallicity. The full treatment of TP-AGB stellar phase in the M05
  models gives significant emission for populations between 0.2 and 1
  Gyr. For similar plots and discussion see M05.}
\label{fig:spectra}
\end{figure}
\section{A challenging issue: the observed rest-frame $K$-Band
  luminosity function at high redshift}
\label{sec:data}

In this paper we focus on a well documented discrepancy between
semi-analytic models and observations, the inability of the
models in matching the observed redshift evolution of the rest-frame
near-infrared galaxy luminosity function \citep{Pozzetti2003,
  Cimatti2004, Kitzbichler2007}. This has recently been confirmed over
a wide redshift range \citep{Cirasuolo2008}. This paper uses
a data set from the Ultra Deep Survey (UDS), the deepest survey from
the UKIRT Infra-Red Deep Sky Survey (UKIDSS), containing imaging in
the $J$- and $K$-bands, with deep multi-wavelength coverage in
$BVRi'z'$ filters in most of the field.  The sample contains $\approx$
50,000 objects over an area of 0.7 square degrees, with high
completeness down to $K\le23$.  \citet{Cirasuolo2008} find that the
space density of the most massive galaxies at high redshifts (above 2)
is under-predicted by semi-analytic models, in other words the
theoretical luminosity function lacks the brightest sources in the
near-IR.


\citet{Tonini2009, Tonini2010, Fontanot2010} showed that the number of
bright $K$-band objects at high redshift in semi-analytic models can
be increased by including the M05 models with their treatment of the
TP-AGB phase of stellar evolution. We briefly recall here the origin
of such an effect. The M05 models, predict that young populations have
a significant contribution to the near-infrared. For the luminosity
function analysis, the differences are expected to be more significant
at high redshift where a larger fraction of the galaxy population
contains young stars. In Fig. \ref{fig:spectra} we plot the spectral
energy distribution (SED) for a population with 1 Gyr of age and solar
metallicity, using M05 and an illustrative example of the stellar
populations previously implemented in semi-analytic models (see M05
for similar plots and discussion). In the $K$-band the M05 model
predicts more than twice the emission, hence affects the prediction of
semi-analytic galaxies at high redshift.



In concluding this section, some words of caution must be given on the
data/model comparison.  In \citet{Cirasuolo2008} the galaxies have
photometric redshifts, which were obtained by fitting empirical as
well as synthetic templates from \citet{Bruzual2003}. For consistency,
photometric redshifts and rest frame magnitudes should have been
derived for the data using the same stellar populations that we are
implementing in the semi-analytic models, but these data are not
available to us.  However we emphasize that the differences that arise
from using the M05 models to convert from mass to light (or light to
mass) are considerably larger than the ones originated from the
determination of photometric redshifts
\citep[e.g.][]{Maraston2006}. The subsequent conversion from observed-
to rest-frame magnitudes is more difficult to track, as the different
theoretical templates usually give different fitted ages depending on
the properties of each galaxy \citep[e.g.][]{Maraston2006,
  Cimatti2008}. However, this will produce differences between derived
k+e corrections that are not systematic and therefore should not alter
our conclusions.

Finally, when comparing model results and data, one should consider
that model magnitudes are ``total'', while observational measurements
are usually based on ``aperture'' magnitudes.  At redshift zero, a
significant fraction of light might be missed for large objects that
exceed the available aperture diameter \citep[e.g.][]{Lauer2007,
  Linden2007}. At the higher redshifts studied here, despite galaxies
being smaller than the maximum available apertures, there can still be
an issue of missing light when small apertures are used to ensure high
signal-to-noise. Moreover, the situation can be complicated by limited
instrumental resolution that might blend together objects in crowded
regions. The first problem is minimized in the data we use by applying
point spread function (PSF) corrections to total magnitudes
\citep{Cirasuolo2008}. Nevertheless, both aspects can influence the
evolution of the bright end of the luminosity function.



\begin{figure*}
\centering
\includegraphics[width=17.9cm]{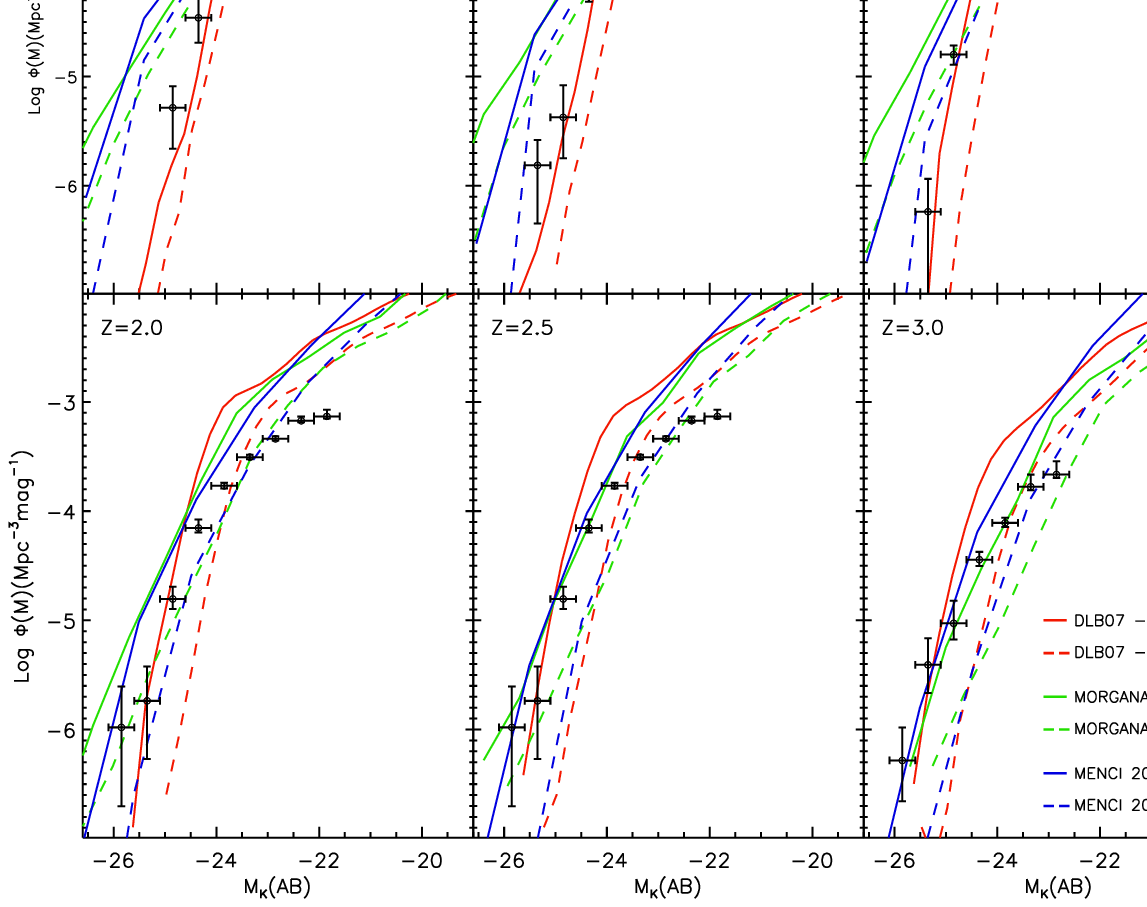}
\caption{The evolution of the K-band luminosity function from z=3.0 to
  z=0.5. The original predictions from three different semi-analytic
  models are shown as dashed lines and their version with the M05 as
  solid lines. Red lines refer to \citet{Delucia2007}, green to
  \citet{Monaco2007} and blue to \citet{Menci2006}. Data from
  \citet{Cirasuolo2008} are shown as black opened circles and the
  solid line.}
\label{fig:kevoall}
\end{figure*}

\section{Results}
\label{sec:results}

\subsection{The $K$-band Luminosity Function}

Fig. \ref{fig:kevoall} compares the evolution of the $K$-band
luminosity function from redshift 3 to redshift 0.5 for the
semi-analytic models with the \citet{Cirasuolo2008} data (shown as
open black circles). \citet{Delucia2007} models are shown in red, {\sc
  morgana} in green and \citet{Menci2006} in blue.  Original model
versions are shown as dashed lines and the M05 versions as solid lines.

The three galaxy formation models in the M05 versions show an enhanced
$K$-band emission (between 0.25 and 0.5 mags) from the brightest
objects ($M_K\le-24$) which for $z\ge2$ brings the models into
agreement with data. The original versions of the models predict that
only old populations provide substantial $K$-band emission. For this
reason, the bright end of the $K$-band luminosity function could only
be built-up at lower redshifts, when old populations become dominant
in massive galaxies. The TP-AGB phase gives a simple and
straightforward way to solve the problem with the observed evolution
of the $K$-band.

The agreement between the semi-analytic-plus-M05 models for bright
$K$-band objects and observations at high redshift is remarkable.  In
principle effects from the age/metallicity degeneracy could produce an
artificial agreement between data and model. For example, a luminous
$K$-band can originate from very metal-rich populations or from much
older populations than those dominated by the TP-AGB
emission. However, the wide redshift range that is spanned by these
observations and the trend of the observed $K$-band LF with redshift
allows us to exclude such effects acting at all redshifts. This is
particularly the case at high redshift, where the time elapsed since
the Big Bang is short enough such that this age degeneracy cannot
enter the game.

These results suggest that, if masses and ages were estimated from
observational data using M05, these would be in agreement with model
predictions.
The conversion to photometric properties was fully responsible for the
disagreement with observations that was pointed out by
\citet{Cirasuolo2008} and previously found by other authors
\citep{Pozzetti2003, Cimatti2004, Kitzbichler2007}. This result has
important implications for the observational determinations of stellar
masses and ages from photometric data, in particular for galaxies at
high redshift. Significant $K$-band emission can be produced by young
populations at high redshift through the TP-AGB stellar phase. Without
considering it, large $K$-band emission can only originate at older
ages, which results in a systematic over-estimation of stellar masses
derived from emission in this band \citep[e.g.][]{Maraston2006}.

Interesting differences among the models emerge at low redshifts.  The
\citet{Delucia2007} model plus M05 follows the bright tail in every
redshift bin (z=0.5 to z=3.0). At redshift 1 the better match between
data and the models implies a certain fraction of $~1$~Gyr population
in these galaxies, a prediction that could be tested by acquiring
rest-frame near-IR spectra.

For {\sc morgana} and \citet{Menci2006}, the inclusion of the M05
models worsens an existing discrepancy at lower redshift $z<=1.5$,
namely that the models over-predict the number density of massive
galaxies. This discrepancy is emphasized with M05 because existing
$\sim$~1 Gyr populations have a higher flux. However, the excess is
present for both versions of each model, hence it is primarily caused
by mass growth rather than age. In both models, this is caused by a
combination of enhanced merging times and inefficient AGN feedback at
low redshift (see Section \ref{sec:models}).

The dynamical friction merging times in \citet{Menci2006} and {\sc
  morgana} are shorter than what is expected from the numerical
analysis of \citet{BoylanKolchin2008}, as pointed out by
\citet{Delucia2010}. Moreover, both models have AGN feedback
implementations that are inefficient in shutting down star
formation in massive objects at later times. In the \citet{Menci2006}
model this is caused by the absence of ``radio-mode'' AGN feedback.
In {\sc morgana} some amount of star formation is required to
destabilize the reservoir of gas and trigger this feedback mode. This
results in galaxy optical colours that are too blue \citep[as pointed
out by][for {\sc morgana}]{Kimm2009} and in an excess of massive
galaxies as emphasised by both stellar population models in our study.

\begin{figure}
\centering
\includegraphics[width=8.8cm]{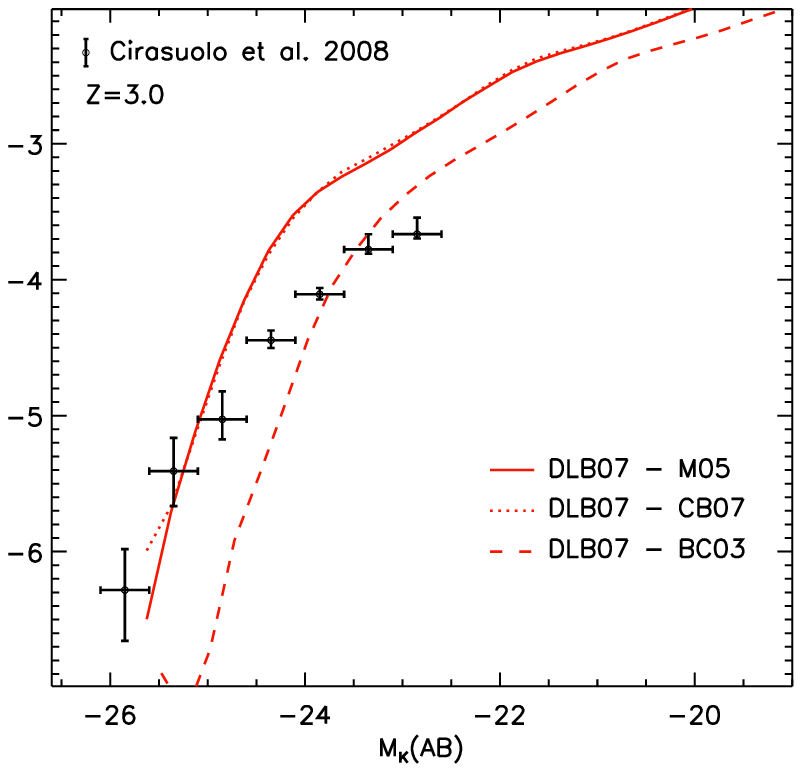}
\caption{The $K$-band luminosity function for the Munich model at
  z=3.0. The solid, dotted and dashed red lines represent,
  respectively, runs using the M05, \citet{Bruzual2007} and
  \citet{Bruzual2003} stellar populations. }
\label{fig:cb07}
\end{figure}

In Fig. \ref{fig:cb07} we plot the $K$-band Luminosity function at
redshift 3.0 for the Munich Model using the original
\citet{Bruzual2003} stellar populations (red dashed line), the M05
(solid red line) and the \citet{Bruzual2007} (dotted red line)
models. The similarity between the M05 and \citet{Bruzual2007} results
shows that the different treatment of the TP-AGB phase (respectively
\citet{Maraston1998} and \citet{Marigo2008}) has a minor impact in our
conclusions.


  
%

\subsection{The Ages of the Galaxy Populations}

\begin{figure*}
\centering
\includegraphics[width=17.9cm]{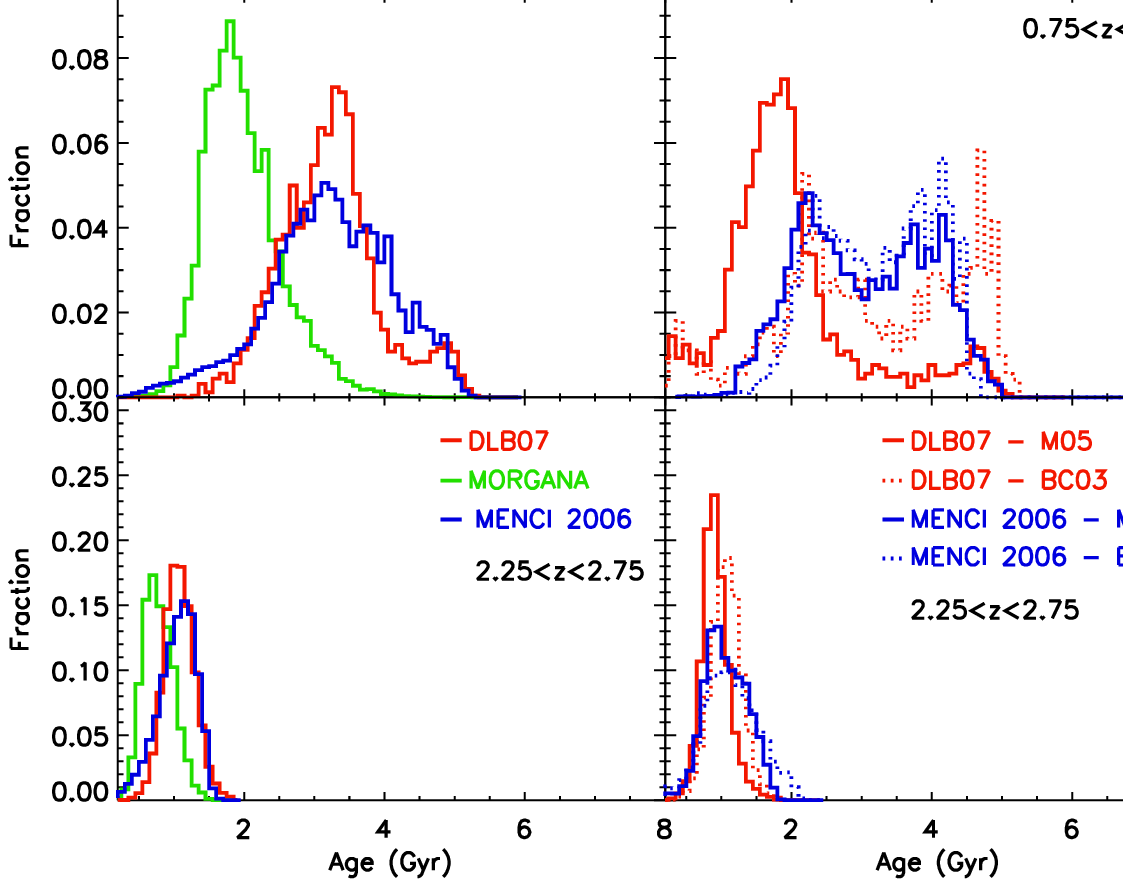}
\caption{The mass-weighted (left-hand panel) and $K$-band light
  weighted (right-hand panel) age distributions of model galaxies with
  M$_K<-24$. Red, green and blue colours refer to \citet{Delucia2007},
  {\sc morgana} and \citet{Menci2006}, respectively.  Solid lines
  represent model versions with M05, whereas dotted lines refer to the
  original stellar populations. From top to bottom, panels refer to
  redshift z=0.5, z=1.0 and z=2.5.}
\label{fig:massage}
\end{figure*}

In order to clarify the different results obtained with the three
semi-analytic models, we show in Fig. \ref{fig:massage} the
mass-weighted and $K$-band light-weighted ages (left-hand and
right-hand panels, respectively) for galaxies brighter than M$_K$=-24,
as a function of redshift. The mass-weighted ages illustrate the
relative contribution of galaxies with different ages to the total
mass budget (with red, green and blue lines representing the Munich,
{\sc morgana} and the \citet{Menci2006} models, respectively).  At
early times (z$\approx$2.5, bottom left-hand panel), the impact of the
TP-AGB phase is larger, because the mean galaxy ages in the three
models are around 1 Gyr. As we move towards lower redshifts (middle
and top left-hand panels) a bimodality emerges in the Munich model,
with the bright-end of the $K$-band luminosity function being built up
by a combination of young and old populations (with the latter,
maintained by the ``radio-mode'' AGN feedback, growing in importance
as we move to lower redshifts).  This bimodality is also present in
\citet{Menci2006}, but is weaker and the oldest population is $\sim$ 1
Gyr younger than in the Munich model. On the other hand, {\sc morgana}
shows considerably younger ages, centered at $\sim$2.5 Gyr, with only
a very weak peak at $\sim$ 4 Gyr. Despite the difference in the age
distribution at z=0.5 (top left-hand panel), the ongoing
star formation produces the same mass excess in both {\sc morgana} and
\citet{Menci2006} (resulting in a similar over-estimation of the
number density of bright $K$-band objects). The \citet{Menci2006}
model has a smaller fraction of younger ages. This is due to the
assumed modeling of the ``quasar-mode'' feedback combined with the
absence of ``radio-mode''. At very high redshift (z$>$3) star
formation is high in the progenitors of massive galaxies due to merger
induced starbursts. At low redshifts only a fraction of these galaxies
have their star formation quenched. On the other hand, {\sc morgana}
has continuous on-going star formation but always at a moderate level,
being self-regulated by the ``radio-mode'' feedback.

The mass-weighted age distribution of the models helped us
understanding the results on the $K$-band luminosity function and
display the backbone of the models. We now consider a light-weighted
age distribution, which emphasises how the different input stellar
population models can force such distributions to different age
domains.  We consider $K$-band light-weighted ages, because they
emphasise the distinction between the model ingredients. For technical
reasons, we cannot easily compute light-weighted ages in the {\sc
  morgana} model. We therefore limit this test to the two other
models.  In the three right-hand panels of Fig. \ref{fig:massage}, red
lines represent results from the Munich model while blue lines give
ages for \citet{Menci2006}. Solid lines represent versions that
include the TP-AGB phase with the M05 and dotted lines the original
stellar populations.
   
The strongest difference between the two model renditions is displayed
by the Munich model, because of the age bimodality mentioned above.
Focussing on redshift 1 and 0.5, the difference between light-weighted
and mass weighted ages is smaller for M05 than for BC03 (solid vs
dotted lines), because the young populations coloured with the M05 get
enough $K$-light such that the mass weighted histogram relative
weights are maintained. In case of the version with BC03, instead, the
$K$-band light only comes from old populations, and as a result the
weight of the bimodality ($k$-light-weighted) is distorted, with a
much higher fraction of populations getting old ages. The behaviour is
diluted for the \citet{Menci2006} models because galaxy ages do not
show a clear bimodality between young and old ages. Therefore,
mass-weighted and $K$-band light-weighted age distributions are
similar and the results for the different stellar population models
are similar as well.

\begin{figure}
\centering
\includegraphics[width=8.8cm]{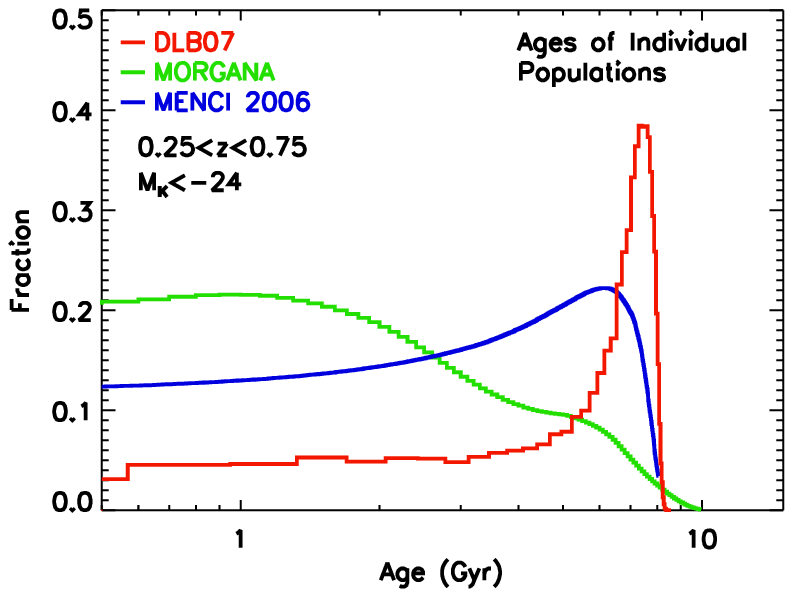}
\caption{The ages of the individual populations present in
  semi-analytic galaxies with M$_K < -24$.  The solid red line
  represents \citet{Delucia2007}, the solid green line shows {\sc
    morgana} predictions and the solid blue line gives
  \citet{Menci2006} ages at z=0.5. }
\label{fig:massage2}
\end{figure}

It should be noted that the mass-weighted age is an average over the
individual populations that compose the theoretical galaxies. This
implies that individual ages can extend down to much lower
values. This can be seen in Fig. \ref{fig:massage2}, where we show the
ages of the individual populations for galaxies with M$_K<-24$ at
z=0.5 in the three semi-analytic models. The 1 Gyr old populations
present in the three models explain the impact of the TP-AGB phase
even at this redshift.  It can also be seen that {\sc morgana}
exhibits the larger fraction of young stars, since the ``radio-mode''
feedback is regulated by star formation. \citet{Menci2006} ages are
considerably older, due to the impact of the ``quasar-mode'' feedback
at early times.  However, as it starts being ineffective at lower
redshifts a considerable fraction of younger populations emerge.

Despite the significant improvement obtained in matching the evolution
of the bright end of the $K$-band luminosity function, the faint end
remains problematic.  The luminosity (as well as the stellar mass)
function for faint objects ($-22\le M_K\le-24$) is known to be much
higher than measured (\citealt{Fontana2006, Weinmann2006b,
  Henriques2008, Fontanot2009b}b) and the inclusion of the M05 models
worsens the case.  This excess can be removed in different ways at
redshift zero, by using a more up-to-date cosmology
\citep{Somerville2008} or combining the disruption of stellar material
from satellites during mergers \citep{Monaco2006, Henriques2008,
  Somerville2008} with more efficient supernova feedback
\citep{Henriques2010, Guo2010}. Nevertheless, the comparison presented
in our work for high redshift (as already shown by
\citealt{Fontanot2009b}b), shows that for the early phases of galaxy
evolution this might be a problem, even considering problems of
incompleteness with the high redshift data.

\section{Summary}
\label{sec:conclusions}

The main objective of this work is to re-address the fundamental
question of matching the observed rest-frame $K$-band luminosity
function of galaxies over the Hubble time, using semi-analytic
models. In the literature \citep{Pozzetti2003, Cimatti2004,
  Kitzbichler2007, Cirasuolo2008}, it has been pointed out that
semi-analytic models underestimate the rest-frame $K$-band galaxy
luminosity of the brightest objects at high redshift ($\sim~2-3$), and
the failure has been attributed to an insufficient mass build-up at
early epochs.

However, the galaxy luminosity function does not only depend on the
mass build-up, but also on the light emitted per unit mass. Hence, in
order to pin down the origin of the mismatch, we improve upon the
rest-frame $K$-band emission from the model galaxies. We use the M05
stellar population models, which include the full treatment of the
emission from the cool and luminous TP-AGB phase. The contribution of
this phase of stellar evolution in the M05 models is important at
intermediate ages (between ~0.2 and 2 Gyrs), which are expected to be
predominant at 2$<$z$<$3.
The relevance of this ingredient has been recently shown for the
semi-analytic model GALICS in \citet{Tonini2009, Tonini2010}, where
the observed near-IR colours of redshift 2 galaxies could only be
matched by the model inclusive of the TP-AGB emission. Similarly,
\citet{Fontanot2010} showed that the inclusion of this stellar phase
in MORGANA increases significantly the number density of EROs at high
redshift.

We consider several semi-analytic models - namely \citet{Delucia2007},
{\sc morgana} \citep{Monaco2007} and \citet{Menci2006} and we
implement the M05 stellar population models, keeping all other
ingredients and assumptions unchanged.

We find that the semi-analytic models with the M05 models exhibit a
brighter $K$-band LF by as much as 0.5 mags at the highest redshift
bins. This is precisely the offset that was plaguing the comparison
with the UKIDSS data for the brightest objects in
\citet{Cirasuolo2008}. Models and data at high redshift and for
M$_K<-24$ now match very well. This result is confirmed when
  using the \citet{Bruzual2007} models as input for the Munich
  model. This confirms that different modelling of the TP-AGB phase
  has a minor impact on our conclusions.

This result is strongly suggestive that the models at redshift $2-3$
do not underestimate mass, rather they did require a proper conversion
between mass and light. Moreover, we show that a strong evolution in
mass predicted by hierarchical models is compatible with no evolution
on the bright-end of the $K$-band LF from z=3 to the local universe.
This means that, at high redshifts and contrary to what is commonly
accepted, $K$-band emission is not necessarily a good tracer of galaxy
mass.

At lower redshift, the details of the implementation of AGN feedback
and merging timescales produce differences between the various
semi-analytic models that are not altered by the inclusion of the M05
models. In particular, {\sc morgana} and \citet{Menci2006} exhibit a
$K$-band luminosity function bright tail that is higher than the data,
which is due to an excessive mass build-up connected to the lack of an
efficient quenching of low-z cooling flows via "radio-mode" feedback.

Similarly, the faint end of the galaxy luminosity function remains
substantially overestimated by the models at all redshifts. This is a
well documented problem (\citealt{Fontana2006, Weinmann2006b,
  Henriques2008, Henriques2010, Fontanot2009b}b; \citealt{Guo2010})
that we plan to study in future work.

In recent years, our understanding of the various phases of stellar
evolution has improved. Moreover, we now have high-quality
observational data covering a wide spectral range (including the
rest-frame near infra-red). Therefore, we should now be able to
constrain galaxy formation models with better accuracy and disentangle
between different theoretical approaches.



\section*{Acknowledgements}

This project is supported by the Marie Curie Excellence Team Grant
MEXT-CT-2006-042754 "UniMass" (PI: C. MAraston) of the Training and
Mobility of Researchers programme financed by the European
Community. PM and FF acknowledge support by the ASI/COFIS grant. FF
acknowledges the support of an INAF-OATs fellowship granted on 'Basic
Research' funds.  GDL acknowledges financial support from the European
Research Council under the European Community’s Seventh Framework
Programme (FP7/2007-2013)/ERC grant agreement n. 202781.

The authors thank Emanuele Daddi and the referee Gustavo Bruzual for
helpful comments. PM and FF thank Laura Silva for useful discussions.
BH, CM and CT thanks Edd Edmondson for a computer network that always
works. BH thanks Peter Thomas for his guidance and constant support.

\bibliographystyle{mn2e}
\bibliography{paper}

\label{lastpage}

\end{document}